\documentstyle[12pt,epsf]{article}
\begin{document}

\newcommand{\newc}{\newcommand}
\newc{\etal}{{\it et al.}\ }
\newcommand{\be}{\begin{equation}}
\newcommand{\ra}{\rightarrow}
\newcommand{\ee}{\end{equation}}
\newcommand{\nn}{\nonumber}
\newcommand{\rpv}{\mbox{$\rlap{\kern0.25em/}R_p$}}
\newcommand{\rp}{\mbox{$R_p$}}
\renewcommand{\l}{\lambda}
\renewcommand{\mark}{{\huge\bf \P\P\P}}

\newcommand{\met}{\mbox{$\rlap{\kern0.25em/}E_T$}}
\renewcommand{\topfraction}{1}
\renewcommand{\bottomfraction}{1}

\begin{titlepage}
\begin{flushright}
{\tt hep-ph/9807547}
\end{flushright}
\vspace{2cm}
\begin{center}
{\Large  \bf
Could we learn more about HERA high $Q^2$ anomaly \\
from LEP200 and TEVATRON? \\[2mm]
$R$-parity violation scenario}

\vspace{1cm}
{\bf A.S.Belyaev} $^{\mbox{a,b,}}$\footnote{e-mail:
belyaev@ift.unesp.br},
{\bf A.V.Gladyshev} $^{\mbox{c,}}$\footnote{e-mail:
gladysh@thsun1.jinr.ru}

\vspace{1cm}
$^{\mbox{a}}$ {\it Instituto de F\'{\i}sica Te\'orica,
              Universidade Estadual Paulista, \\
              Rua Pamplona 145, 01405--900 S\~ao Paulo, Brazil}\\
$^{\mbox{b}}$ {\it Skobeltsin Institute for Nuclear Physics,
Moscow State University, \\ 119 899, Moscow, RUSSIA}

\vspace{.5cm}

$^{\mbox{c}}$ {\it Bogoliubov Laboratory of Theoretical Physics,
     Joint Institute for Nuclear Research, \\
	141 980 Dubna, Moscow Region, RUSSIA}

\end{center}

\vspace{1cm}

\begin{abstract}
The excess of high $Q^2$ events at HERA reported in the early 1997 by
H1 and ZEUS collaborations has become the subject
of extensive studies in the framework of several models related to new
physics.  Here we  concentrate  on the most promising, from our point
of view, model describing HERA anomaly.  We update our previous
analysis~\cite{first} and take into account new HERA statistics of the
1997 year.  HERA events are considered within the $R$-parity broken
SUSY model for a specific scenario with several non-zero couplings.
$R$-parity broken SUSY with several non-zero couplings could explain both
high $Q^2$ $e^+ + jets$ and $\mu^+ + jets$ observed at HERA. The
consequence of such a particular scenario is the excess of high $Q^2$
di- or tri-jet events at HERA.  The relation of this scenario for LEP
and TEVATRON colliders is considered.  This study shows that if a
squark resonance does take place at HERA, supersymmetry with
broken $R$-parity can be  revealed at either LEP200 or TEVATRON in the
near future.  \end{abstract}

\end{titlepage}

%%%%%%%%%%%%%%%%%%%%%%%%%%%%%%%%%%%%%%%%%%%%%%%%%%%%%%%%%%%%%

\section{Introduction}

Anomalous high-$Q^2$ events observed at HERA by H1~\cite{h1} and
ZEUS~\cite{zeus} collaborations in the early 1997 have provoked much
efforts of their explanation within different theoretical frameworks.
For the 1994-96 year data an excess of high $Q^2$ events was at about
3$\sigma$ level.  Later, these two collaborations reported an analysis
with the 1997 year data included at the Symposium on Lepton-Photon
Interactions~\cite{lph} and very recently at DESY seminar on the 13th of
March, 1998~\cite{desy-seminar}.

The present statistics is more than doubled (46.6 pb$^{-1}$ - ZEUS and
37.04 pb$^{-1}$ - H1) in comparison with the statistics of the 1994-96
years (20.1 pb$^{-1}$ - ZEUS and 14.2 pb$^{-1}$ - H1). The acceptance
has also been increased.  At present, for the 1994-97 data combined,
the excess is at the $2\sigma$ level, for example, for the
above-mentioned luminosity the number of observed events with $Q^2 >
20000$ GeV$^2$ is 7 (ZEUS) + 10 (H1) = 17,  while the number of
predicted events is only 5.3 (ZEUS) + 4.4 (H1) = 9.7

Basically there are three kinds of proposed explanations: \\
1) contact-term interactions of very heavy bosons, heavy
leptoquark-type objects being exchanged between a positron and a
quark~\cite{cont}; \\
2) non-standard evolution of the parton densities~\cite{part}; \\
3) leptoquark~\cite{lq} or squark~\cite{sq} resonance within the
$R$-parity violated SUSY model.

The first two scenarios, as it has been shown, do not give quite a
good explanation of HERA events~\cite{alt-status}.

We concentrate here on the possibility of the resonant  $e^+ q$
production at HERA which gives the most reasonable explanation of high
$Q^2$ events.  One can expect resonance production of LQ from
valence ( $u$ and $d$) or sea ($\bar{u}, \bar{d}, s$)  quarks.

However, the resonance production of $e^+\bar{u}$ and $e^+\bar{d}$ is
excluded by HERA data for the $e^- p$ mode with 1 pb$^{-1}$ statistics,
because such signals would be seen in the $e^- u$ and $e^- d$ resonance
production  for quite a big coupling constant value of an order of
0.2~\cite{ellis} appearing in the scenario with a resonance in $e^+p$
collisions.

On the other hand, the generally accepted leptoquark framework of
Ref.~\cite{lq-frame} with a non-zero coupling only within one family is
excluded for HERA by the combined CDF+D\O \  limit on the leptoquark
mass of 240 GeV at 95\% C.L.~\cite{lq-lim}.  This limit is established
under the assumption that $BR(LQ\rightarrow e+q)=1$.  But for HERA,
within this model, the branching ratio is fixed to unity if one takes
into account the severe limit ($5\times 10^{-7}$) on the product of
chiral couplings from the branching ratio of the leptonic pion
decay~\cite{davidson}.

That is why we  would like to consider a more preferable, from our
point of view scenario, when leptoquark appears as a scalar
superpartner of a quark within the $R$-parity violating (RPV) SUSY
model~\cite{rpv}.

We would also like to update our previous paper~\cite{first} with the
1997 HERA data included, namely, we consider a possibility of the
$R$-parity violation provided $s$-channel resonance at HERA for the
scenario with several non-zero $R$-parity violating couplings takes
place.

The starting point for our analysis is the superpotential of the
Minimal Supersymmetric Standard Model with $R$-violating terms included
\begin{eqnarray}
W &=&  h_E L E^c H_1 + h_D Q D^c H_1 + h_U Q U^c H_2 +
\mu H_1 H_2  \nn \\
&& + \mu_i H_2 L_i+
\l_{ijk} L_i L_j E^c_k +
\l'_{ijk} L_i Q_j D^c_k +
\l''_{ijk} U^c_i D^c_j D^c_k
\end{eqnarray}

Couplings $\l_{ijk}$ and $\l''_{ijk}$ have the following
properties:  $\l_{ijk}=-\l_{jik}$, $\l''_{ijk}=-\l''_{ikj}$.

The terms with $\l$ and $\l'$ violate a lepton number $L$ while the last
term with $\l''$ violates a baryon number $B$. The $B$ and $L$ violating
operators, if they exist simultaneously, lead to a fast proton
decay~\cite{tata}.  Thus, only one kind of operators is allowed. HERA
events can be explained in the framework of the $R$-parity broken SUSY
model only if $\l'\ne 0$. That is why we hereafter consider the case
with $\l''=0$.

The  $\l$ and $\l'$ -terms result in the following Lagrangian in
component fields:
\begin{eqnarray}
L_{\rpv,\l}&=&\l_{ijk}[\tilde{\nu}_{iL}\bar{e}_{kR}e_{jL}+
\tilde{e}_{jL}\bar{e}_{kR}\nu_{iL}+
{\tilde{e}^*}_{kL}\overline{(\nu_{iL})^C}e_{jL} -
(i\leftrightarrow j)]_F + h.c.,\\
L_{\rpv,\l'}&=&\l'_{ijk}[\tilde{\nu}_{iL}\bar{d}_{kR}d_{jL}+
\tilde{d}_{jL}\bar{d}_{kR}\nu_{iL}+
{\tilde{d}^*}_{kL}\overline{(\nu_{iL})^C}d_{jL} - \nn \\
&&\tilde{e}_{iL}\bar{d}_{kR}u_{jL}-\tilde{u}_{jL}\bar{d}_{kR}e_{iL}-
{\tilde{d}^*}_{kL}\overline{(e_{iL})^C}u_{jL}]
 + h.c.
\end{eqnarray}

The strictest limits on different single  couplings
are presented in Table~\ref{tab:bounds} (see,
{\it e.g.}~\cite{dreiner,bhat} and references therein).

\begin{table}[htb]
\begin{center}
\begin{tabular}{|cc||cc|cc|cc|}\hline
$ijk$ & $\l_{ijk}$ & $ijk$ & $\l'_{ijk}$ & $ijk$ & $\l'_{ijk}$ &
$ijk$ & $\l'_{ijk}$ \\ \hline
121 & $0.05^{a\dagger}$ & 111 & $0.001^d$& 211 & $0.09^h$ & 311 &
$0.16^k$ \\
122 & $0.05^{a\dagger}$ & 112 & $0.02^{a\dagger}$ & 212 & $0.09^h$ &
312& $0.16^k$ \\
123 & $0.05^{a\dagger}$ & 113 & $0.02^{a\dagger}$ & 213 & $0.09^h$ &
313& $0.16^k$ \\
131 & $0.06^b$  & 121 & $0.035^{e\dagger}$ & 221 & $0.18^i$ & 321 &
$0.20^{f*}$ \\
132 & $0.06^b$  & 122 & $0.06^c$ & 222 & $0.18^i$ & 322& $0.20^{f* }$ \\
133 & $0.004^c$  & 123 & $0.20^{f*}$ & 223 & $0.18^i$ & 323& $0.20^{f*}$ \\
231 & $0.06^b$ & 131 & $0.035^{e\dagger}$ & 231 & $0.22^{j\dagger}$ &
331& $0.26^g$ \\
232 & $0.06^b$ & 132 & $0.33^g$ & 232 & $0.39^g$ & 332& $0.26^g$ \\
233 & $0.06^b$ & 133 & $0.002^c$& 233 & $0.39^g$ & 333& $0.26^g$ \\
\hline
\end{tabular}
\end{center}
\caption{The strictest experimental bounds on $R$-parity violating
couplings for the squark mass ${\tilde m}=100$ GeV}
\label{tab:bounds}
\end{table}

The bounds denoted by a dagger $(^\dagger)$ are $2\sigma$ bounds, the
other bounds are at the $1\sigma$ level. The bounds denoted by
an asterisk $(^{*})$ are based on the assumption of absolute mixing
in the (SM) quark sector~\cite{dreiner}.

The bounds in Table \ref{tab:bounds} are obtained from the following
processes: \\
${(a)}$ charged current universality \cite{bgh,pdg}, \\
${(b)}$ $\Gamma(\tau\ra e\nu{\bar\nu})/\Gamma(\tau\ra\mu\nu{\bar\nu})$
\cite{bgh,pdg}, \\
${(c)}$ bound on the mass of $\nu_{e}$ \cite{hallsuzuki,tata,wark}, \\
${(d)}$ neutrinoless double-beta decay \cite{moha,klapdor}, \\
${(e)}$ atomic parity violation \cite{davidson,wood,atomic}, \\
${(f)}$ $D^0-{\bar D}^0$ mixing \cite{dreiner,gupta,agashe}, \\
${(g)}$ $R_\ell =\Gamma_{had}(Z^0)/\Gamma_\ell ( Z^0)$
\cite{srid2,bhatt}, \\
${(h)}$ $\Gamma(\pi\ra e{\bar\nu})/\Gamma(\pi\ra\mu{\bar\nu})$
\cite{bgh}, \\
${(i)}$
$BR(D^+\ra{\bar K}^{0*}\mu^+\nu_\mu)/BR(D^+\ra{\bar K}^{0*}e^+\nu_\mu)$
\cite{bhattchoud,bhatt}, \\
${(j)}$ $\nu_{\mu}$ deep-inelastic scattering \cite{bgh}, \\
${(k)}$ $BR(\tau\ra\pi\nu_\tau)$ \cite{bhattchoud,bhatt}, \\
${(l)}$ heavy nucleon decay \cite{sher}, \\
${(m)}$ $n-{\bar n}$ oscillations \cite{zwirner,sher}.

Bounds shown in Table~\ref{tab:bounds}  scale with the following
factors:  \\
$a)$, $b)$, $e)$, $g)$ to $m)$  --  $M_{\tilde{q,l}} / 100 \ GeV$; \\
$c)$, $f)$ -- $\sqrt{M_{\tilde{q,l}} / 100 \ GeV}$; \\
$d)$ --  $(M_{\tilde{q}}/100\ GeV)^2\sqrt{M_{\tilde{g}}/1\ TeV}$.

Within the model under study only the following processes survive
if one takes into account the above-mentioned low energy limits:
\begin{equation}
e^+d\rightarrow \tilde{t}, \;\;
e^+d\rightarrow \tilde{c}, \;\;
e^+s\rightarrow \tilde{t}
\end{equation}

We would like to concentrate ourselves on the scenario with only a
valent quark involved.  The branching ratio of stop (scharm) decay into
an electron and a quark should be less than unity in order not to
contradict D\O \ and CDF data.  There are two ways of having this: \\
1) squark has also  $R$-parity conserving decay channels; \\
2) squark has no $R$-parity conserving decay channels but has
{\it additional $R$-parity violating decay channels}, which leads in
turn to the flavor changing neutral current (FCNC). The $R$-parity
conserving decay channels will be closed if,
$m_{\tilde{t}}< m_b+m_{\tilde{W_1}}$ and
$m_{\tilde{t}}< m_W+m_{\tilde{b}}$.  The same situation could be for
$c$-squark but for a smaller SUSY parameter space~\cite{ellis}.

The  first possibility has been  intensively studied (see, for example,
Ref.~\cite{ellis,dp}).  The $R$-parity conserving decay will lead to the
charged current event excess.  Actually, HERA has some  excess of such
events~\cite{lph}. Since the observed charged current events are events
with a single jet and $R$-parity conserving  decays lead to multiparton
final states, a special assumption is required that some  partons
should be too soft to be detected as a jet.

The second possibility to  have $BR(\tilde{q}\ra e^+,d)<1$ has been
considered in our previous paper~\cite{first}.  We considered the case
of  several non-zero couplings of the same order $\simeq$ 0.04.
However, when one considers such a scenario, one should also take into
account limits from the low energy experiments on the products of
$R$-parity violating couplings. Almost all combinations of several
non-zero couplings contributing to the flavor changing neutral current
are excluded.  Modern limits on the products of $R$-parity violating
couplings come  from: \\
1) $\nu-e$  conversion in nuclei~\cite{prod:nu-e}; \\
2) limits from  rare leptonic decays of the long-lived neutral kaon,
muon,tau and from mixing of neutral K- and B-mesons~\cite{prod:kb}; \\
3) from $B^0$ decays to two charged leptons~\cite{prod:b0dec}; \\
4) from muon decay to electron and photon~\cite{prod:mudec};\\
5) from features of $K-$ and $B-$  systems~\cite{prod:kbsyst}.

In Table~\ref{tab:prodlim} we present the most important for our
analysis bounds from the literature cited above.
\begin{table}[htb]
\caption{\label{tab:prodlim}}
\begin{center}
\begin{tabular}{|l | l || l | l  |}
\hline
combination & upper bound &combination & upper bound \\
\hline
\hline
$|\l_{121}\l'_{211}|$ & $4.0\cdot 10^{-7} \cite{prod:nu-e}$ &
$|\l_{131}\l'_{211}|$ & $1.0\cdot 10^{-5} \cite{prod:nu-e}$
\\
$|\l_{121}\l'_{131} | $ &$ 4.6\cdot 10^{-5} \cite{prod:b0dec}$ &
$|\l_{131}\l'_{131} | $ &$ 4.9\cdot 10^{-4} \cite{prod:b0dec}$
\\
$|\l_{121}\l'_{213} | $ &$ 4.6\cdot 10^{-5} \cite{prod:b0dec}$ &
$|\l_{131}\l'_{231} | $ &$ 4.6\cdot 10^{-4} \cite{prod:b0dec}$
\\
$|\l'_{121}\l'_{222}| $ & $3.5\cdot 10^{-8} \cite{prod:kb}$ &
$|\l'_{131}\l'_{232}|$  & $3.5\cdot 10^{-7} \cite{prod:kb}$
\\
$|\l_{132}\l'_{131} | $ &$ 6.0\cdot 10^{-4} \cite{prod:b0dec}$ &
$|\l_{122}\l'_{131} | $ & $2.4\cdot 10^{-5} \cite{prod:b0dec}$
\\
$|\l'_{121}\l'_{221}|$ & $8.0\cdot 10^{-8} \cite{prod:nu-e}$ &
$|\l'_{131}\l'_{221}|$ & $2.0\cdot 10^{-6} \cite{prod:nu-e}$
\\
$|\l'_{131}\l'_{231}|$ & $2.0\cdot 10^{-6} \cite{prod:nu-e}$ &
$|\l'_{121}\l'_{231}|$ & $8.0\cdot 10^{-8} \cite{prod:nu-e}$
\\
$|\l'_{j2n}\l'_{i2n}|$ & $4.8\cdot 10^{-5} \cite{prod:kb}$ &
$|\l'_{in2}\l'_{jn1}|$ & $4.8\cdot 10^{-5} \cite{prod:kb}$
\\
$|\l'_{121}\l'_{221}| $ &$ 8.1\cdot 10^{-4} \cite{prod:mudec}$ &
$|\l_{131}\l_{231}  | $ &$ 2.0\cdot 10^{-4} \cite{prod:mudec}$
\\
$|\l'_{131}\l'_{132}| $ &$ 7.7\cdot 10^{-4} \cite{prod:kbsyst}$ &
$|\l'_{331}\l'_{332}| $ &$ 7.7\cdot 10^{-4} \cite{prod:kbsyst}$
\\
$|\l'_{311}\l'_{322}| $ &$ 6.1\cdot 10^{-6} \cite{prod:kbsyst}$ &
$|\l'_{311}\l'_{332}| $ &$ 6.1\cdot 10^{-6} \cite{prod:kbsyst}$
\\
$|\l'_{331}\l'_{333}| $ &$ 6.1\cdot 10^{-6} \cite{prod:kbsyst}$ &
$|\l'_{131}\l'_{123}| $ &$ 1.4\cdot 10^{-6} \cite{prod:kbsyst}$ \\
\hline
\end{tabular}
\end{center}
\end{table}

After careful analysis of the relevant bounds only the following
combinations  of non-zero $\l'$ survive:\\
1) for the $c$-squark resonance scenario:
$\l'_{121}, \l'_{323}, \l'_{333}, \l'_{311}$
and $\l_{131}$;\\
2) for the $t$-squark resonance scenario:
$\l'_{131}, \l'_{333}, \l'_{311}$
and $\l_{131}$.
In addition to these non-zero couplings there is a possibility
for $\l'_{112}$ $or$ $\l'_{113}$ to be non-zero as well. We will
consider this case in relation to possible signals of $R$-parity
violation at LEP.

We chose non-zero $\l'$'s equal to each other. An interesting fact is
that bounds on the products of  different $\l'$ close the decay channel
both for $\tilde{t}$ and $\tilde{c}$ into a muon and a quark and the
only possibility for an additional channel of squark decay into $\tau$
and b-quark is left
($|\l'_{121}\l'_{222}|,|\l'_{131}\l'_{232}|<3.5\cdot 10^{-7}$,
\cite{prod:kb}).

The purpose of the present paper is to consider this scenario with
several  non-zero $R$-parity violating couplings for the explanation of
the excess of high $Q^2$ neutral current DIS events at HERA  as well as
FCNC events observed by H1 and discuss a possible check of such a
scenario at LEP200 and TEVATRON.

All analytic and most of the numerical calculations have been
performed with the help of the CompHEP software
package~\cite{comphep}.This package allows one to make complete
tree level calculations in the framework of any fed model. We have
implemented into this package the part of the supersymmetric Standard
Model with $R$-parity violating terms  relevant to our analysis.

\section{HERA events}

To describe HERA anomalous events, we used '94-97 data. For example,
for $Q^2>20000$ GeV$^2$ we have 17 events while only 9 are predicted by
the SM.  We have also taken into account the integrated luminosity of
ZEUS+H1 equal to $83.6$ pb$^{-1}$ and an average detector efficiency
0.85.

\begin{figure}[htb]
  \vspace*{-1.2cm}
  \begin{center}
    \leavevmode
    \epsfxsize=14cm
    \epsffile{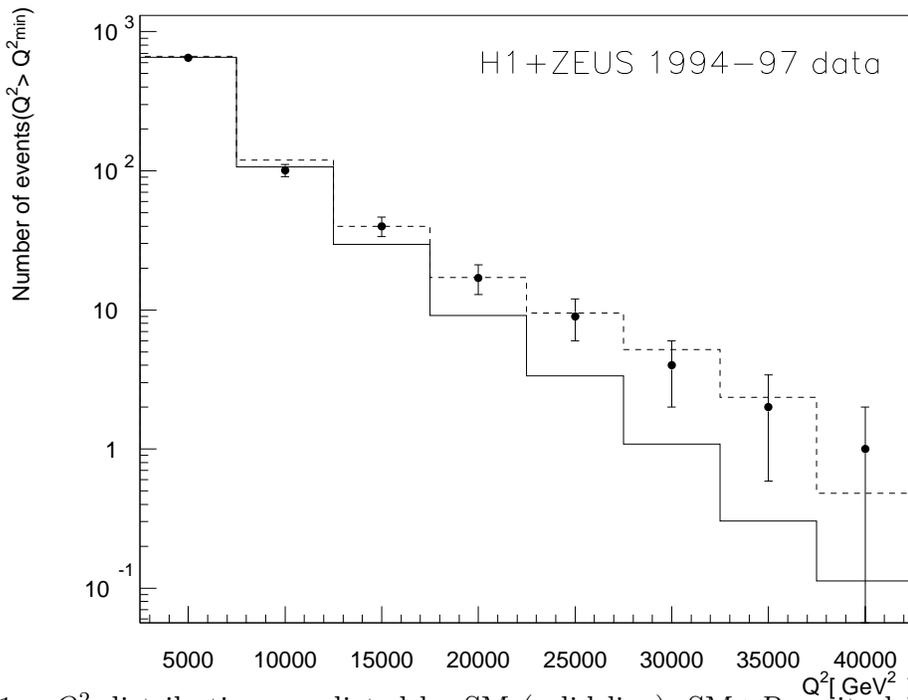}
  \vspace*{-1.8cm}
     \caption{
	$Q^2$ distribution, predicted by SM (solid line),
     SM+$R$-parity breaking squark resonance (dashed line)
     [$\l ' =0.038$] and  data (solid dots).\label{heraq2}}
\end{center}
  \vspace*{1.0cm}
\end{figure}

Distribution for $Q^2>Q^2_{min}$ for the H1+ZEUS 1994-97 data and that
for the   SM prediction is shown are Fig.~\ref{heraq2}. One can see that
the systematic deviation from the SM starts from $Q^2>15000$ GeV$^2$,
which is shown in Table~\ref{tab:numev}.

\begin{table}[htb]
\caption{\label{tab:numev}}
\begin{center}
\begin{tabular}{ | l | l | l | l | l | l | l | l | l |}
\hline
$Q^2>Q^2_{min}$ [GeV$^2]$& 5000 & 10000 & 15000 & 20000& 25000&
30000 & 35000& 40000 \\
\hline
DATA   & 762 & 117   & 42   & 17  & 9   & 4   & 2    & 1 \\
SM     & 731 & 115.1 & 31.7 & 9.7 & 3.6 & 1.2 & 0.48 & 0.2 \\
\hline
\end{tabular}
\end{center}
\end{table}

Combined systematic and statistical errors for a number of events,
predicted by the SM are of an order of 10\%.

In Table~\ref{tab:heram} we present values for invariant mass
of the events with $Q^2>20000$~GeV$^2$.

\begin{table}[htb]
\caption{\label{tab:heram}}
\begin{center}
\begin{tabular}{ | l | l | l | l | l | l | l | l | l | l | l | }
\hline
H1 & 1 & 2 & 3 & 4 & 5 & 6 & 7& 8 & 9 & 10  \\
\hline
MASS  & 245(*)& 217& 210 & 208 & 206 & 201(*)& 200(*) & 180(*)
& 173 & 157 \\
\hline
\hline
ZEUS& 1 & 2 & 3 & 4 & 5 & 6 & 7&&&  \\
\hline
MASS   & 252& 228& 226(*) & 207 & 194(*) & 192& 170  &&& \\
\hline
\end{tabular}
\end{center}
\end{table}

The events from the '97 year data are marked by an asterisk (*). We
have taken mass values from the figure of~\cite{desy-seminar}.

One can see that the events observed by the H1 collaboration
($Q^2>20000$~GeV$^2$) are more concentrated around the invariant mass of
200 GeV than the ZEUS events.  In conclusion of PH'97 it was said that
ZEUS and H1 data did not fit well together for the scenario with a
single narrow resonance~\cite{lph}.  But the statistics and mass
resolution are not enough at the moment to exclude completely the
scenario with one resonance: for $Q^2>20000$, 7 H1 events and 3 ZEUS
events are in the mass window $200\pm 20$ GeV.  Moreover, one can
obtain a broad mass distribution if a superposition of two close mass
eigenstates is considered~\cite{t1t2-mix}.

In our calculations we have used the CTEQ3M structure
function~\cite{cteq} and $Q^2$ scale equal to $(p_e - p'_e)^2$. The
total cross section describing the observed data turned out to be
0.30~pb\footnote{ We obtained this value using data presented by H1 and
ZEUS collaborations~\cite{lph,desy-seminar}. This value is the same as
was presented at 'LP97:  the statistics has been increased since that
time and 3 new events for $Q^2 > 20000$ GeV$^2$ were observed by H1
collaboration.}, while the Standard Model gives only 0.16~pb.

We use the set of non-zero  $R$-parity violating Yukawa couplings,
which has been described above. We use:\\
1)$\l'_{121}=\l'_{323}=\l'_{333}=\l'_{311}=\l'$, $\l_{131}=\l$\\
or\\
2)$\l'_{131}=\l'_{333}=\l'_{311}=\l'$, $\l_{131}=\l$.

We have calculated the complete set of Feynman diagrams for the process
$e^+ + p \to e^+ + jet + X$ in the framework of the $R$-broken
supersymmetric model for HERA energy.

\begin{figure}[htb]
  \vspace*{-1.5cm}
  \hspace*{3.0cm}
 \begin{center}
    \leavevmode
    \epsfxsize=15cm
    \epsffile{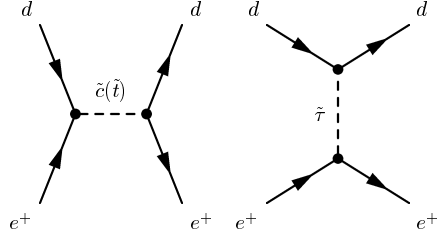}
    \vspace*{-16cm}
    \caption{\label{diag-hera}
	Diagrams for the $e^+ +p \to e^+ + jet +X$ process in the
     supersymmetric model with $R$-parity violation.}
  \end{center}
\end{figure}

The relevant Feynman diagrams are shown in
Fig.\ref{diag-hera}.

As it is naturally expected, the main contribution
comes from the diagrams with a resonant squark in the $s$-channel
(99\%).  Contribution from the $t$-channel diagram is negligible (1\%)
for the case under consideration.  However, these diagrams give the
same order contribution as a {\it nonresonant} $s$-channel  diagram
with a heavy sparticle, but this is not the subject of our study.

In our calculations we assumed NLO corrections by introducing the
$K$-factor equal to 1.25 for $m_{\tilde q}=200$ GeV~\cite{mike}.

The presence of several non-zero couplings leads to the branching ratio
$BR(\tilde c(\tilde t) \to e^+ d)$ equal to 50\%.

We obtained the  value of $\l'$ equal  to 0.038 which describes the
event excess in HERA data with the 200 GeV $s$-channel squark resonance.
Fig.~\ref{heraq2} shows the $Q^2$ distribution for the Standard Model
and SUSY model with broken $R$-parity tuned to describe data and the
combined ZEUS+H1 data distribution itself.

One can see that the $Q^2$ shape of deviation from SM is well
described by the $s$-channel squark resonance.

The presence of another non-zero coupling --- $\l'_{323}$
or $\l'_{323}$ causes the flavour changing neutral current events:
$e^+ d\rightarrow \tau b$. \\
Leptonic and hadronic $\tau$-decay channels lead to the
following signatures:\\
1) $e+$ or  $\mu^+ + $\mbox{$\rlap{\kern0.25em/}P_T$}$ + jet$;\\
2) di- or tri-jet events + $\mbox{$\rlap{\kern0.25em/}P_T$}$.

Two outstanding FCNC events with a high $p_T$ $\mu^+ + jet(s) + missing
\; P_T$ signature have been observed by H1 collaboration and reported
at Jerusalem conference~\cite{jerus}.  Only 0.05 events were predicted
by the SM in the kinematic region of these two events.  The scenario
with non-zero couplings describes well the kinematics of the observed
events.  If one takes into account the
$\tau\rightarrow\mu\bar{\nu_\mu}\nu_\tau$ decay branching ratio equal
to 17\%, then one will have about 2 $\mu^+ +jet + missing \; P_T$
events which were observed by H1 collaboration.   If
the deviation from the SM for di- or tri-jet +
$\mbox{$\rlap{\kern0.25em/}P_T$}$ events
is found, it will be an additional favour for the scenario with several
non-zero RPV couplings.

\section{Implication for LEP200 and TEVATRON}

Having in mind that processes with $R$-parity violation could occur in
$e^+ e^-$ collisions at LEP, we consider the possibility of detecting
it in the pair jet production via a sparticle exchange. Deviation of
the jet production rate from the SM at LEP  due to $R$-parity violating
SUSY has been the subject of several papers~\cite{rpv1lep,first,rpv2lep}.

\begin{figure}[thb]
  \vspace*{-1.0cm}
 \hspace*{2.0cm}
  \begin{center}
    \leavevmode
    \epsfxsize=15cm
    \epsffile{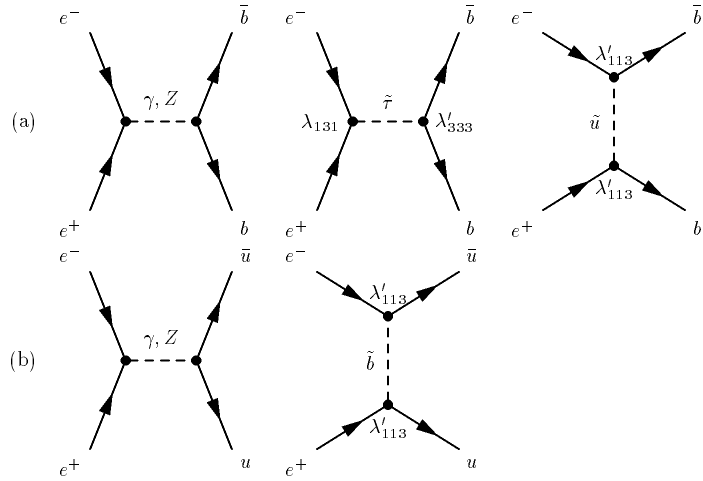}
    \vspace*{-12.0cm}
    \caption{\label{diag-lep}
	Diagrams for the $e^+ e^-\to q\bar{q}$ process
     in the $R$-parity violated SUSY model.}
  \end{center}
\end{figure}

Comparing with the Standard Model, some excess in dijet production is
expected. For this purpose, we have calculated the complete set of
relevant diagrams presented in Fig.\ref{diag-lep}. We consider the case
with also non-zero $\l'_{113} = \l'$.  Hereafter, we assume that
sleptons have a mass 200 GeV unless otherwise noted.  We also assume
that sleptons are not produced resonancely, considering as an
example the scenario with $\sqrt{s}_{e^+e^-}$ = 184 GeV.

\begin{figure}[thb]
  \vspace*{-1.0cm}
  \begin{center}
    \leavevmode
    \epsfxsize=8cm
    \epsffile{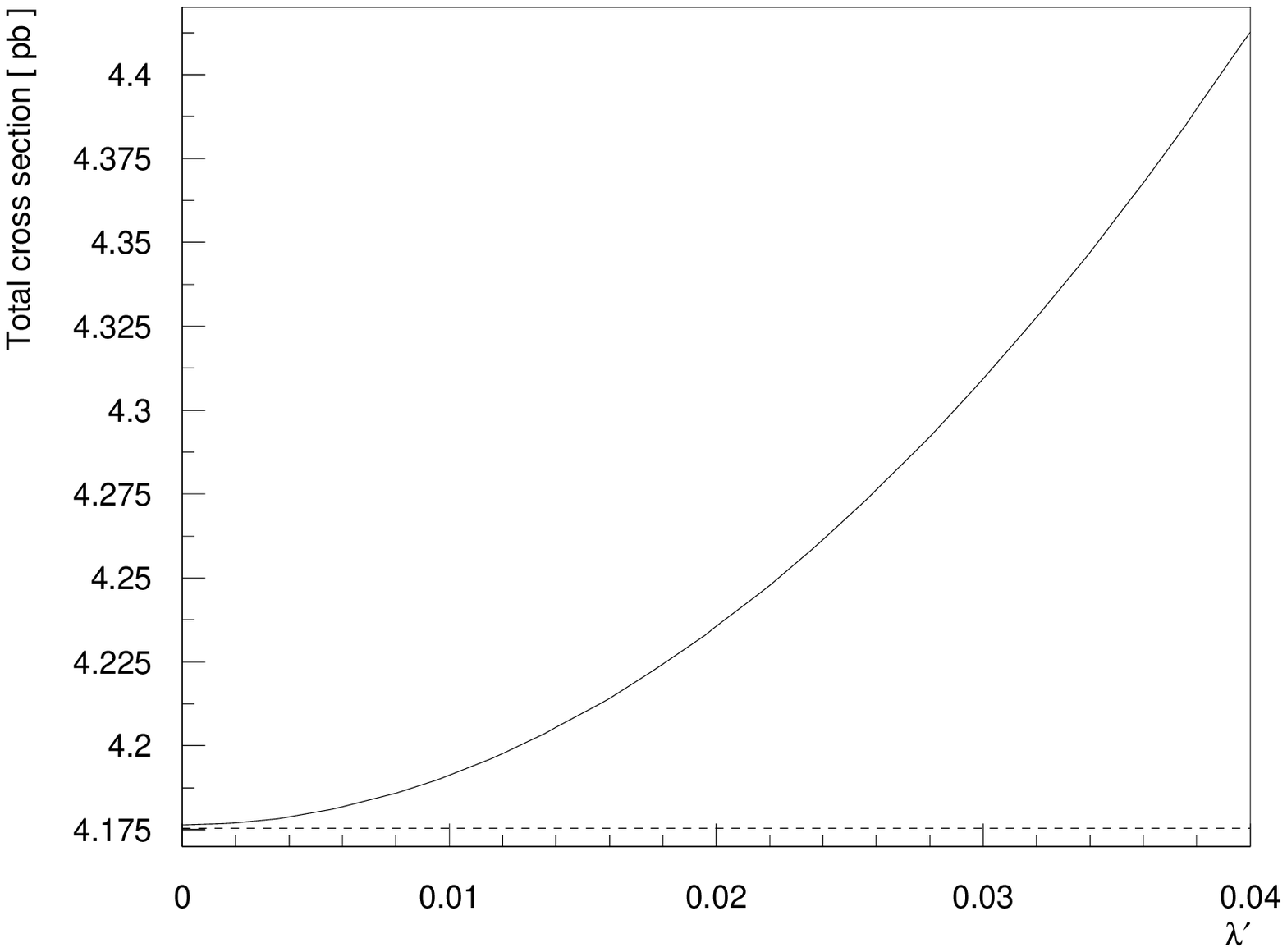}
    \put(-170,140){a)}
    \put(-150,140){$e^+ e^- \to b\bar{b}$}
    \put(-150,125){$\sqrt{s}=184$~GeV}
    \put(  60,80){b)}
    \put(  80,80){$e^+ e^- \to u\bar{u}$}
    \put(  80,65){$\sqrt{s}=184$~GeV}
    \epsfxsize=8cm
    \epsffile{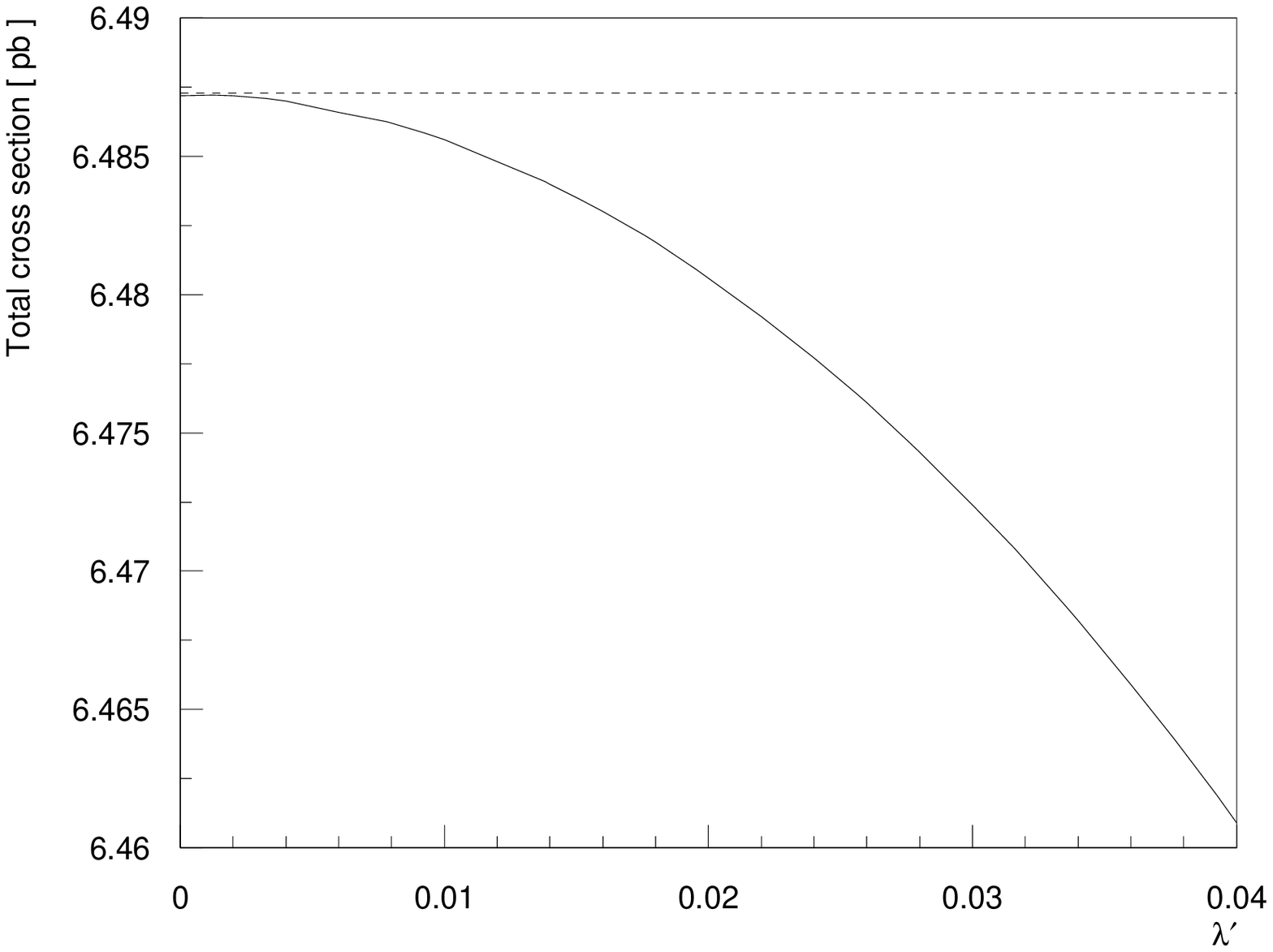}
    \vspace*{-1cm}
    \caption{\label{lep}
	Total cross section of $e^+ e^-\to q\bar{q}$
    versus $\l'$ for $b$-quark (a) and $c$-quark (b) pair production
    at LEP. Dashed line --- the SM prediction.}
  \end{center}
\end{figure}

We would like to stress that there is a difference (which was missed in
some earlier papers) in the up and down quark production at LEP for the
scenario under study.  For bottom-quark production, for example, there
is an $s$-channel diagram with a sneutrino and $t$-channel
diagram with an $u$-squark (Fig.\ref{diag-lep} (a)).  It makes the
total cross section dependent not only on $\l'$ but also on  $\l$
values. The $t$-channel diagram has negative interference while the
$s$-channel diagram has a positive contribution to the total cross
section.  The total effect from both the diagrams is positive, when
$\l'$ and $\l$ are of the same order of magnitude.  At the same time,
if only the $t$-channel diagram with $R$-parity violating vertex is
present, as it happens for the up-quark pair production
(Fig.\ref{diag-lep} (b)), the total effect is negative due to the
negative $t$-channel diagram interference and the cross section is
smaller than the SM one.  This is illustrated in Fig.\ref{lep} for the
$b$-quark pair production (a) and the $c$-quark pair production (b),
where the total cross section as a function of  $\l'$ is shown ($\l$ is
fixed equal to 0.038).

If we take $\l_{131}$ equal to its upper limit (0.12), we have a
$5\%$ deviation ($\simeq 0.21$~pb) of the total cross section from the
Standard Model for $\sqrt{s}=184$~GeV LEP. In our calculation we put
the value of $\l'$ to that describing the HERA data, namely, 0.038.  If
we assume that the b-jet production ratio can be measured with the 1\%
accuracy, one can establish  the limit on $\l$  about 0.055 ($\l'$
= 0.038) or the limit on the product $|\l\cdot\l'|<2\cdot 10^{-3}$.

As for the up-quark production, the $t$-channel diagrams bring a  small
negative contribution.  One could expect some contribution of a 200 GeV
SUSY particle to the $R_b$ value very accurately measured at LEP90
which is equal to $0.2170\pm 0.0009$~\cite{ewdata}. But it turns out
that for $\sqrt{s}=90$~GeV the contribution from additional diagrams to
the $b\bar{b}$ production is only in the 6-th digit.

Though limits on $\l$'s are not so strict, it seems unnatural if their
values are much higher than $\l'$.  We would like to look at the
case when $\l$ is small or equal to 0 and the excess in the $b\bar{b}$
production is absent.
For example, for $\l\simeq\l'=0.038$ the $b\bar{b}$ production
deviates from the Standard Model by less than  1\%.  By a lucky
chance we have at present  three different  colliders  with energy of
an order of 1 TeV:  lepton-lepton --- LEP, lepton-hadron --- HERA and
hadron-hadron --- TEVATRON. These machines are highly complementary to
each other. We show below that possible failure to find the $R$-parity
broken SUSY for our scenario
(if it really exists) at LEP leads to the luck at
TEVATRON in this search and vice versa.
One of the most interesting effects for TEVATRON revealing
the $R$-parity violation is the single top production via a
slepton decay~\cite{snglrpv1,first,snglrpv2} which is resonancely produced
in $p\bar{p}$ collisions.

\begin{figure}[thb]
  \vspace*{-1.5cm}
 \hspace*{3.0cm}
  \begin{center}
    \leavevmode
    \epsfxsize=15cm
    \epsffile{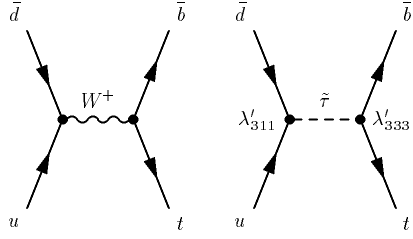}
    \vspace*{-16.0cm}
    \caption{\label{diag-tev}
	Diagrams for $2\to2$ single top production
     processes at TEVATRON in the $R$-parity violated SUSY model.}
  \end{center}
\end{figure}

The  Feynman diagrams for the $2 \to 2$ process of
single top production are shown in Fig.~\ref{diag-tev}.

\begin{figure}[thb]
  \vspace*{-2.0cm}
 \hspace*{2.0cm}
  \begin{center}
    \leavevmode
    \epsfxsize=15cm
    \epsffile{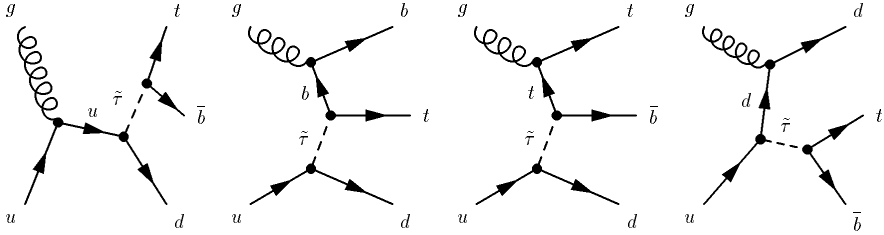}
    \vspace*{-16.0cm}
    \caption{\label{diag-tevwg}
	$\tilde\tau$-gluon fusion and resonant $\tilde\tau$ $2\ra 3$
	diagrams for single top production
     processes at TEVATRON in the $R$-parity violated SUSY model.}
  \end{center}
\end{figure}

In comparison with our previous paper we take into account also
$2\ra 3$ diagrams of single top production via $R$-parity violating
vertices:  $\tilde\tau$-gluon fusion process and $2\ra 3$ process with
resonant $\tilde\tau$ production. The corresponding diagrams are
presented in Fig.~\ref{diag-tevwg}.

\begin{figure}[thb]
  \vspace*{-1.0cm}
  \begin{center}
    \leavevmode
    \epsfxsize=8cm
    \epsffile{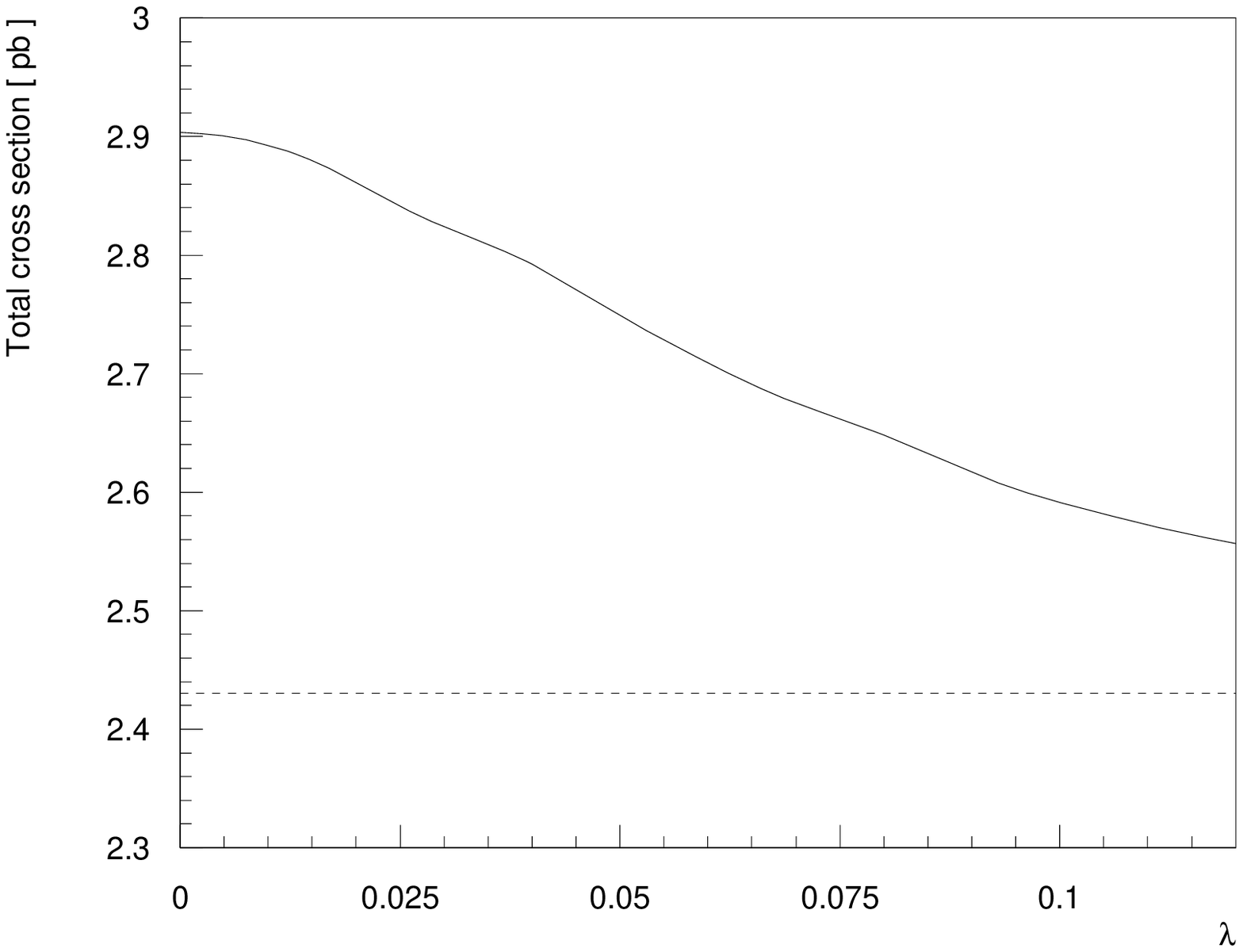}
    \put(-130,150){a)}
    \put(-110,150){$p \bar{p}' \to t\bar{b} (q) + X$}
    \put(-110,135){$\sqrt{s}=1.8$~TeV}
    \put(  70,150){b)}
    \put(  90,150){$e^+ e^- \to b\bar{b}$}
    \put(  90,135){$\sqrt{s}=184$~GeV}
    \epsfxsize=8cm
    \epsffile{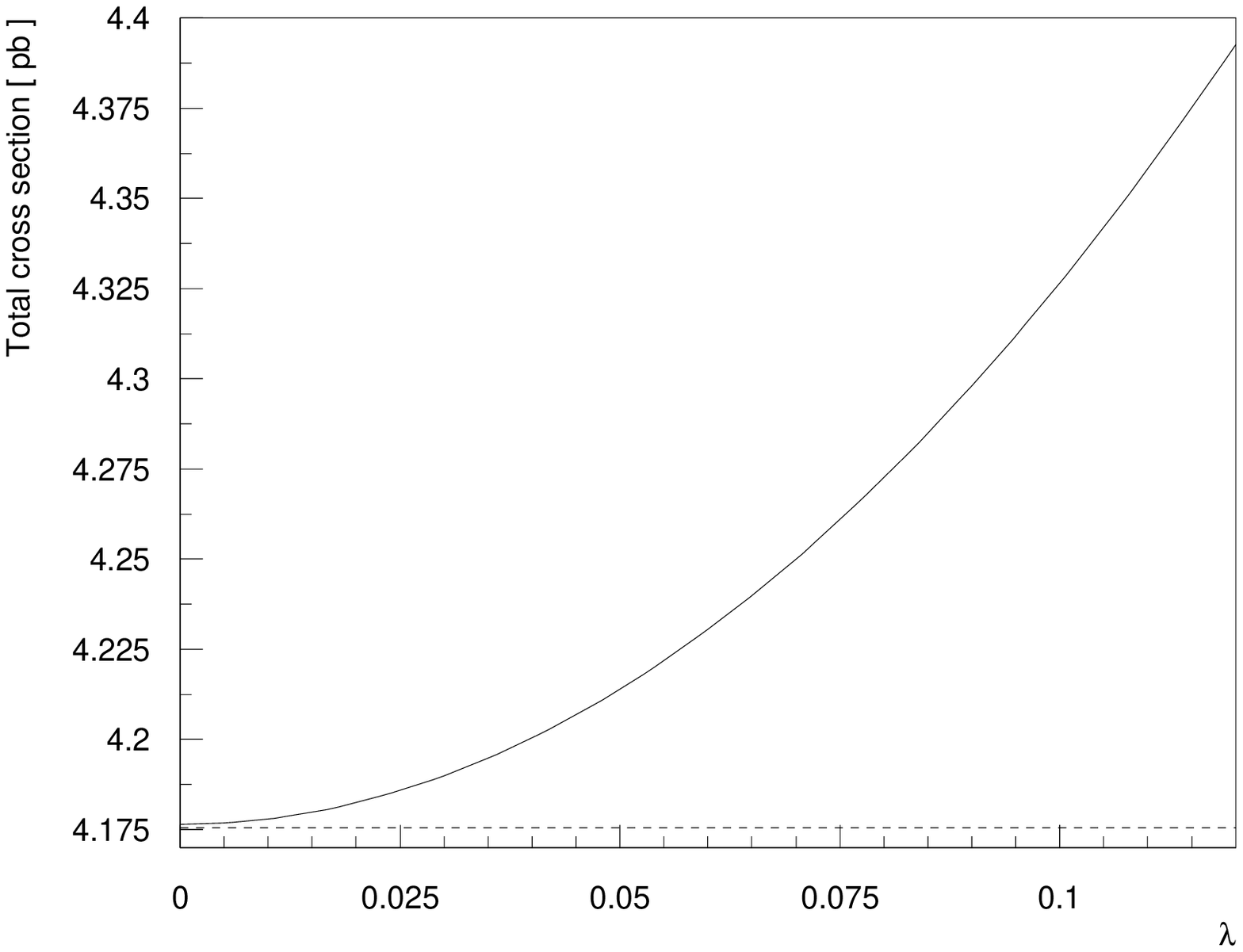}
  \vspace*{-1.0cm}
    \caption{\label{tev}
	Total cross section of the $p\bar{p}\to t\bar{b}(q)$
	single top production process versus $\l$ (a) and total cross
	section of the $b\bar{b}$ production at LEP versus $\l$ (b).
	Dashed line shows the SM prediction ($l'_{311}=\l'_{333}=0.038$)}
\end{center}
\end{figure}

In spite of the fact that the cross section is directly proportional
only to $\l'$, there is also dependence of the cross section on the
$\l$ coupling, which is related to the width and branching ratios of the
selectron decay.  For comparatively large $\l$ (0.1-0.2)
$\tilde\tau$ mainly decays leptonically and the single top production
is suppressed. In this case, the deviation in the $b\bar{b}$ production
could be measured at LEP. On the other hand, if  $\l$ is small or even
equal to 0, there is a significant contribution to the single top
production via processes mentioned above.

In our calculations we have fixed $\l'_{311}$ and $\l'_{333}$ equal to
0.038 and studied the dependence of the single top quark cross section
on the values of $\l_{131}$.  We assume $m_{top}$=175 GeV, use the
CTEQ3M structure function~\cite{cteq} and use the slepton mass as a QCD
scale.  We have calculated the following processes:\\
1) $u\bar{d}\ra t\bar{b}, d\bar{u}\ra \bar{t}b;$\\
2) $ u g\ra t\bar{b}d, dg\ra \bar{t}bu,
\bar{u} g\ra \bar{t}b\bar{d}, \bar{d} g\ra t\bar{b}u$.\\
The process $2\ra 3$ has about 20\% contribution to the total
RPV cross section and thus should be taken into account.  For example,
for $\l_{131}$=0, the RPV cross section of the $2\ra 2$ process is 0.32
pb, while the RPV cross section of $2\ra 3$ is 0.078 pb.

For  our  analysis we have used  NLO cross sections for the SM single
top production using the results of Ref.\cite{st-nlo}($m_t$=175 GeV,
$\sqrt{s}$=2 TeV): $\sigma(t\bar{b})$=$0.88\pm 0.05$,
$\sigma(Wg=tq\bar{b}+tq)= 2.43 \pm 0.4$.  NLO cross section is about
20\% higher than the LO one (for details see~\cite{our-st}).  One could
expect the same order corrections for the RPV diagrams of the single top
quark production.  We used this factor in our analysis.

We can see that the contribution of the RPV processes can be as large
as 20\% of the total single top production rate.  It is  illustrated in
fig~\ref{tev} where the dependence of this deviation versus $\l$ ($\l'$
is fixed to 0.038) is shown.  We include the results of the analysis for
single top  background processes into our study (see~\cite{moriond}). A
careful kinematic analysis allows one to suppress significantly the
background and makes signal/background ratio equal to
1/2-1/3~\cite{moriond} which in turn gives the possibility to measure
the single top production cross section with the accuracy 19\% and 7\%
for TEVATRON Run2 and Run3, respectively~\cite{our-st}. At the same
time, the top- and bottom-quark invariant mass distribution analysis
will be more sensitive to the RPV single top physics than the total
cross section measurement.

We have found that if one takes into account the 10\% accuracy of the
single cross section measurement, the lower limit on  $\l_{131}$
can be established: $\l_{131}> 0.07$   ($\l'=0.038$). On the other
hand, LEP data will allow one to put an upper limit on $\l_{131}$:
$\l_{131}< 0.055$ ($\l'=0.038$).  We can see that by lucky coincidence
LEP and TEVATRON complement each other in search for supersymmetry with
broken $R$-parity.  The existence of the RPV physics will be checked at
LEP and TEVATRON which together with HERA could prove the validity, or
restrict, or even exclude the scenario with several non-zero RPV Yukawa
couplings.

\section*{Conclusions}

The excess of high $Q^2$ HERA events can be reasonably explained within
the Supersymmetric Standard Model with broken $R$-parity.

New  1997 year data have decreased the statistical significance up to
$2\sigma$ level. It is by no means conclusive and one should wait for
new coming statistics.

We have suggested some specific scenario for the $R$-parity violating
SUSY which has several non-zero couplings.  This scenario could be a
reasonable explanation of two outstanding FCNC muon events observed by
H1 collaboration.  At the same time, one can expect an excess of di- or
tri-jet events.  We have
checked that the Yukawa couplings (as well as their products)
responsible for processes under study do not exceed their experimental
bounds. Processes with $R$-parity breaking can also take place in $e^+
e^-$ collisions at LEP200 and TEVATRON, which are complementary to each
other. If  $R$-parity violation really takes place,  as has been shown
above, it can be  revealed at either TEVATRON or LEP in the near
future.

\vspace{1cm}

We would like to thank D.I.Kazakov and W. de Boer for valuable
discussions and stimulation of this study. Our work has been supported
by the Russian Foundation for Basic Research, grants \# 96-02-17379,
\#96-02-19773-a  and ICFPM  1996 grant. A.S.B. would like to thank
O.Eboli, S.Novaes, R.Rosenfeld and Y.Sirois for fruitful discussions.
He is also grateful to the Instituto de F\'{\i}sica Te\'orica for
kind hospitality.  This work was supported by Funda\c{c}\~ao de Amparo
\`a Pesquisa do Estado de S\~ao Paulo (FAPESP).  A.V.G. acknowledges
the hospitality of the FERMILAB Theory Division, where the final stage
of the work was done.

\end{document}